\documentclass[12pt]{article}

\usepackage{amsmath,amssymb,amsthm,epsf,epsfig, graphicx,latexsym}

\newcommand{\n}{\noindent}

\newcommand{\be}{\begin{equation}}
\newcommand{\ee}{\end{equation}}
\newcommand{\bea}{\begin{eqnarray}}
\newcommand{\eea}{\end{eqnarray}}
\newcommand{\bml}{\begin{subequations}}
\newcommand{\eml}{\end{subequations}}
\newcommand{\bfig}{\begin{figure}}
\newcommand{\efig}{\end{figure}}

\newcommand{\bg}{\beta}

\newcommand{\ve}{\varepsilon}

\newcommand{\lb}{\lambda}
\newcommand{\sg}{\sigma}

\begin{document}

\begin{center}
{\bf {\Large Generalized particle dynamics: modifying
the motion of particles and branes}}\\
\vspace{5ex}

Sudipta Das$^{a,}$\footnote{E-mail: sudipta\_jumaths@yahoo.co.in},
Subir Ghosh$^{a,}$\footnote{E-mail: sghosh@isical.ac.in},
Jan-Willem van Holten$^{b,}$\footnote{E-mail: t32@nikhef.nl} \\
and Supratik Pal$^{a,}$\footnote{E-mail: supratik\_v@isical.ac.in}
\vspace{3ex}

$^a$Physics and Applied Mathematics Unit, Indian
Statistical Institute, \\
203 B.T.Road, Kolkata 700108, India
\vspace{2ex}

$^b$Nikhef,  PO Box 41882, 1009 DB Amsterdam, Netherlands
\vspace{5ex}

{\bf Abstract}
\end{center}

\n {\small We construct a generalized dynamics for particles
moving in a symmetric space-time, i.e.\ a space-time admitting one
or more Killing vectors. The generalization implies that the
effective mass of particles becomes dynamical. We apply this
generalized dynamics to the motion of test particles in a static,
spherically symmetric metric. A significant consequence of the new
framework is to generate an effective negative pressure on a
cosmological surface whose expansion is manifest by the particle
trajectory via embedding geometry \cite{bwg,embed2,embed,pal}.
This formalism thus may give rise to a source for dark energy in
modelling the late accelerating universe. }


\section{Introduction}

According to Einstein's theory of general relativity (GR) the
motion of test particles in a specific background geometry is
described by geodesics of space-time. However, for particles with
spin and/or charge deviations are expected due to spin-orbit
coupling \cite{regge, rvh} or external fields.

Tests of Einstein's theory are usually limited to special
geometries, such as that of spherical masses (e.g., stars). Some
indications exist that deviations from geodesic motion arises over
long distances and time scales, such as for example the famous
Pioneer anomaly \cite{pioneer}.

In this paper we propose a modification of Einstein's equations
of motion for particles in a fixed symmetric space-time, i.e.\
a space-time admitting one or more Killing vectors. The main
ingredient is a coupling between the constant of motion associated
with a Killing vector and the orbit of the particle, generalizing
the concept of spin-orbit coupling. Like the equation of motion for
particles in flat Minkowski space, the action is only invariant under
special co-ordinate transformations defined by the Killing vectors \cite{holt}.
In the context of GR we therefore expect our prescription to
arise in specific geometries as an effective action, taking into
account special effects from the background geometry.

We apply our formalism to the specific case of Schwarzschild
geometry. We introduce an additional non-minimal coupling between
orbital motion and angular momentum and we make explicit the changes
in the effective potential for test particles arising from our
modification. We show how the conventional results are recovered
in the limit of vanishing coupling.

An interesting cosmological application arises from applying
our new dynamics to brane models of cosmology. Observing that
new physics seems to be required to explain dark energy as the
dominant constituent of our present-day universe, it would be
very interesting as well as convenient if one can understand
the negative pressure behavior - a signature of dark energy -
from a modified brane dynamics. Indeed, the modifications of
dynamics we propose can incorporate this feature in an effective
theory, by {\it{non-minimal}} coupling to conventional gravity.

Taking cue from the braneworld gravity \cite{bwg} which arises
from the embedding of the 4-dimensional world in higher dimension, we
follow closely the thin-shell formalism of \cite{embed1,embed2} where one
develops the Friedman equations for cosmology, governed by the induced
gravity on the brane. In our case we use the effective metric emerging
from the particle back reaction. We show that
{\it{ with this extension
one can have a universe which interpolates between the early decelerating and late time accelerating phases.}}
Indeed, it is satisfying that in the present setup one can visualize the early decelerating and late time
accelerating phases of the universe in a unified framework.

In our analysis of the model we exploit both lagrangian and the
hamiltonian framework, using Dirac's formulation \cite{dir} of
constrained dynamics. Finally the relevance of Schwarzschild
geometry to a wide class of sources, as implied by Birkhoff's
theorem \cite{birkhoff}, motivates our choice of studying this
example in more detail.

\section{The generalized particle formalism}

We propose a generalized action for point particles in a
space-time admitting one or more Killing vectors, taking the form
\begin{equation}
S = \int L d\tau = m\, \int d\tau \left[ \frac{1}{2e} g_{\mu\nu}
\dot{x}^{\mu} \dot{x}^{\nu} - \frac{e}{2} - \lambda g_{\mu\nu}
\xi^{\mu} \dot{x}^{\nu} + \frac{e \lambda^{2}}{2} g_{\mu\nu}
\xi^{\mu} \xi^{\nu} +  \frac{e \beta \lambda^{2}}{2}\right].
\label{action}
\end{equation}
Here $\tau$ is the worldline evolution parameter, $x^{\mu}(\tau)$
are the particle co-ordinates, $e(\tau)$ is the worldline einbein
introduced to make the action reparametrization invariant \cite{brinketal},
and $\lb(\tau)$ is an auxiliary worldline scalar variable. Furthermore $\bg$
is a constant, whilst the metric $g_{\mu\nu}(x)$ and the vector
$\xi^{\mu}(x)$ are functions of the co-ordinates $x^{\mu}$.

The action is invariant under infinitesimal co-ordinate transformations
of the form $\delta x^{\mu} = \alpha \xi^{\mu}(x)$, provided the
Lie-derivative of the metric with respect to $\xi$ vanishes:
\begin{equation}
\xi^{\lambda} \partial_{\lambda} g_{\mu\nu} + g_{\nu\lambda}
\partial_{\mu}\xi^{\lambda} + g_{\mu\lambda} \partial_{\nu}\xi^{\lambda} = 0.
\label{2}
\end{equation}
This shows that $\xi^{\mu}(x)$ is the Killing vector associated
with the symmetry of the metric. If the metric admits more than
one Killing vector, the action (\ref{action}) can be extended accordingly.

In the following we use the notation,
\be
\dot{x}^2 = g_{\mu\nu} \dot{x}^{\mu} \dot{x}^{\nu}, \hspace{2em}
\xi \cdot \dot{x} = g_{\mu\nu} \xi^{\mu} \dot{x}^{\nu}, \hspace{2em}
\xi^2 = g_{\mu\nu} \xi^{\mu} \xi^{\nu}.
\ee
The lagrangian equations of motion for the worldline variables
$(e, \lb)$ then imply
\begin{equation}
\frac{\dot{x}^{2}}{e^{2}} = -1 + \lambda^2(\xi^{2}+\beta)~~~;
~~~\lambda=\frac{1}{e}\frac{\xi \cdot \dot{x}}{\xi^{2}+\beta}.
\label{constraints}
\end{equation}
These equations can be used to eliminate $e$ and $\lb$ from the action,
leading to a classically equivalent expression
\begin{equation}
\tilde{S} = - m \int d\tau\, \sqrt{- \dot{x}^2 +
 \frac{(\xi \cdot \dot{x})^2}{\xi^2 + \bg}}.
\label{sqrtaction}
\end{equation}
It is interesting to note that Eq (2.5) is a somewhat generalized
version of the point particle action derived in \cite{bks} in the
context of noncommutative Snyder geometry \cite{snyder}. The
associated canonical momenta are
\begin{equation}
p_{\mu} = \frac{\partial \tilde{S}}{\partial \dot{x}^{\mu}}
 = \frac{m}{e}\, g_{\mu\nu} \left( \dot{x}^{\nu} - \xi^{\nu}
  \frac{\xi \cdot \dot{x}}{\xi^2 + \bg} \right),
\label{lagrmomentum}
\end{equation}
which satisfy the constraints
\begin{equation}
\xi \cdot p = m \bg \lb, \hspace{2em}
p^2 + \frac{1}{\bg} ( \xi \cdot p)^2 + m^2 = 0.
\label{momconstraint}
\ee
In these last equations $e$ and $\lb$ are to be interpreted as
short-hand notation for the solutions of eqs.\ (\ref{constraints}).
\vspace{2ex}

\n
{\it{Hamiltonian formulation}}\\
To analyze the dynamics implied by the action (\ref{action}),
we follow the hamiltonian analysis of constrained systems as
formulated by Dirac \cite{dir}. The canonical momenta are given by
\begin{equation}
p_{e}=\frac{\partial L}{\partial \dot{e}} = 0~~; ~~
p_{\lambda}=\frac{\partial L}{\partial \dot{\lambda}} = 0~~; ~~
p_{\mu}=\frac{\partial L}{\partial \dot{x^{\mu}}} =
 m g_{\mu \nu} \left( \frac{1}{e}\dot{x^{\nu}}-\lambda \xi^{\nu} \right).
\label{can}
\end{equation}
The Hamiltonian follows,
\begin{equation}
H=p_{e} \dot{e} + p_{\lambda} \dot{\lambda} + p_{\mu} \dot{x^{\mu}} - L =
\frac{e}{2m}[p^{2}+m^{2} + 2 m \lambda(\xi \cdot p) - \bg m^2 \lambda^{2}].
\label{hamilt}
\end{equation}
We have two primary constraints \be \psi_{1} \equiv
p_{e}\approx0~~~;~~~ \psi_{2} \equiv p_{\lambda}\approx 0,
\label{p} \ee leading to two secondary constraints, \be \dot\psi_2
\equiv \psi_{3} \equiv e(\xi \cdot p - m \bg \lambda) \approx 0;
~~ \dot\psi_1 \equiv\psi_{4} \equiv
p^{2}+m^{2}+\frac{1}{\beta}(\xi \cdot p)^{2} \approx 0. \label{s}
\ee Notice that $e$ is not allowed to vanish and hence $\psi_{3}$
can be replaced by $\tilde{\psi}_{3} \equiv \xi \cdot p - m
\lambda \beta \approx 0$. Equivalently, we can remove the
auxiliary pair $e,p_e$ by fixing the gauge $e=1$. The remaining
three constraints are non-commutating in general and we may obtain
the first-class constraint (related to the generator of the
reparametrization invariance) \cite{dir} by appropriate linear
combination of $\psi_2,~\psi_3,~\psi_4$. However, as $\xi^\mu$ is
a Killing vector one can check that $\psi _4$ commutes with itself
as well as with $\psi_2,~\psi_3$. Hence $\psi_4$ is the
first-class constraint. Indeed, $\psi_4$ coincides with the
hamiltonian (\ref{hamilt}) and the lagrangian constraint
(\ref{momconstraint}) upon enforcing the remaining second-class
constraints $\psi_2$ and $\tilde{\psi}_3$. Thus
\begin{equation}
2 m H = \psi_4 = p^{2}+m^{2}+\frac{1}{\beta}(\xi \cdot p)^{2} = 0.
\label{constr}
\end{equation}
Clearly this constraint is a generalization of the normal
mass-shell condition $p^2 + m^2 \approx 0$ one is familiar with.

Of course, we must also deal with the second-class constraints by replacing
the canonical Poisson brackets by Dirac brackets \cite{dir}, but in the
present case this does not affect the canonical symplectic structure of the
remaining variables. As in the conventional case our gauge choice
implies $x^{0}=t$ and we can proceed to the standard hamiltonian dynamics.

\section{An application to Schwarzschild geometry}

We now apply our generalized dynamics to the case of a particle in
a Schwarzschild background, with line element (in natural units $c
= G =1$)
\begin{equation}
g_{\mu\nu} dx^{\mu} dx^{\nu} = - \left(1 - \frac{2M}{r}\right)
dt^2 + \frac{dr^2}{1 - \frac{2M}{r}} + r^2 d\theta^{2} + r^2
\sin^{2}\theta d\phi^{2}.
\end{equation}
As is well-known, the symmetries of the metric lead to four
Killing vectors: one related to time translation (conservation of
energy) and the other three deal with rotational symmetry, i.e.
conservation of angular momentum. Out of these three, one deals
with the magnitude whereas the other two are related to the
direction of angular momentum, which means that the particle will
move in a plane. Let us choose this plane to be the equatorial
plane $z=0$ , or $\theta = \frac{\pi}{2}$. Then we can reduce the
above metric on the equatorial plane to
\begin{equation}
g_{mn}dx^m dx^n = - \left(1 - \frac{2M}{r} \right) dt^2 +
\frac{dr^2}{1 - \frac{2M}{r}} + r^2 d\phi^{2}
\end{equation}
The Killing vectors associated with time translations and with rotations
in the equatorial plane carry over to the reduced metric. In particular,
in this frame the Killing vector associated with angular momentum is
$\xi^{\mu}_{(\phi)} = (0,0,1)$. Incorporating this Killing vector into
our general action (\ref{action}), we have
\begin{equation}
S = \int L d\tau =
 m \int \left[ \frac{1}{2 e}\left( - \left( 1-\frac{2M}{r}\right) \dot{t}^{2}
 + \frac{\dot{r}^{2}}{1-\frac{2 M}{r}}+r^{2} \dot{\phi}^{2}\right)
 - \frac{e}{2} - \lambda r^{2} \dot{\phi} \right. $$$$
 \left. + \frac{e \lambda^{2}}{2}\left( r^{2} + \beta \right) \right] d\tau .
\end{equation}
The  canonical momenta turn out to be
\begin{equation}
p_{e} = \frac{\partial L}{\partial \dot{e}} = 0
~~~; ~~~
p_{\lambda} = \frac{\partial L}{\partial \dot{\lambda}} = 0~~,
\nonumber
\end{equation}

\begin{equation}
p_{r} = \frac{\partial L}{\partial \dot{r}} = \frac{m \dot{r}}{e
\left(1-\frac{2 M}{r}\right)}
~~~; ~~~
p_{t} = \frac{\partial L}{\partial \dot{t}} =
 -\frac{m \dot{t}}{e} \left( 1-\frac{2 M}{r}\right)~~, \nonumber
\end{equation}

\begin{equation}
p_{\phi} = \frac{\partial L}{\partial \dot{\phi}} =
\frac{m}{e}r^{2} \dot{\phi} - m \lambda r^{2}.
\label{n10}
\end{equation}
The Hamiltonian is
\begin{eqnarray}
H &=& p_{e} \dot{e} + p_{\lambda} \dot{\lambda} + p_{r} \dot{r} +
p_{t} \dot{t} + p_{\phi} \dot{\phi} - L \nonumber \\
{} &=& \frac{e}{2m}\left[m^2 - \frac{p_{t}^{2}}{1-\frac{2 M}{r}} +
\left( 1-\frac{2 M}{r}\right) p_{r}^{2}+\frac{p_{\phi}^{2}}{r^{2}}
+ 2 m \lambda p_{\phi} - \beta m^2 \lambda^{2}\right].
\end{eqnarray}\
Performing the same constraint analysis as before whilst fixing the gauge
$e = 1$, we find the modified mass-shell constraint and the relation,
\begin{equation}
 2m H = m^2 - \frac{p_{t}^{2}}{1-\frac{2 M}{r}} +
 \left( 1-\frac{2 M}{r}\right) p_{r}^{2} + \left( \frac{1}{r^{2}}
 + \frac{1}{\beta} \right) p_{\phi}^{2} = 0~~~;~~~
p_{\phi} = m \beta \lambda.
\label{constr1}
\end{equation}
Consequently the Hamiltonian equations of motion give the two conserved
quantities,
\begin{equation}
\dot{p_{\phi}} = \{p_{\phi} , H\} = 0~~,
\end{equation}
\begin{equation}
\dot{p_{t}} = \{p_{t} , H\} = 0.
\end{equation}\
So we have $ p_{\phi} = l $ and $ p_{t} = E $, where $l$ and $E$
are two different constants. It is worth mentioning here that $l$
and $E$ are nothing but the angular momentum and energy
respectively, for the particle moving in Schwarzschild background,
but now depending on the parameter $\beta $ as well. In the large
$\beta $ limit the standard results are regained.

This is the Hamiltonian way of reproducing the property that for any
Killing vector $\xi^\mu$
\begin{equation}
\xi \cdot p = c,
\end{equation}
where $c$ is time independent. For the Killing vector shown before,
\begin{equation}
\xi_{(\phi)} \cdot p = p_{\phi}= l = m \beta \lambda,
\label{l}
\end{equation}
where $l$ is a constant and the last equality follows from the constraint.
From the time-like Killing vector: $\xi^{\mu}_{(t)} = (1,0,0)$ we obtain
energy, the other constant of motion,
\begin{equation}
\xi_{(t)} \cdot p = p_{t} = E.
\label{E}
\end{equation}
These are just the same quantities we have obtained previously by
using the Dirac brackets. Thus, those constants are nothing but
the angular momentum and energy in the generalized particle
dynamics, respectively.


\section{Analysis and interpretation of the Effective Potential}

By substitution of the radial momentum (\ref{n10}) and the constants of
motion (\ref{E}) and (\ref{l}) into the mass-shell condition (\ref{constr1})
we find an expression for the radial velocity:
\begin{equation}
\dot{r}^2 + \frac{2}{m}\, V_{eff}(r) = \ve^2,
\label{}
\end{equation}
with $\ve = E/m$ and
\be
2m V_{eff} = \left( 1 - \frac{2M}{r} \right) \left( m^2 +
\frac{l^2}{\bg} + \frac{l^2}{r^2} \right).
\label{pot}
\ee
In the limit $\bg \rightarrow \infty$ this reduces to the standard effective
potential for particles in a Schwarzschild background.

Following standard practice, we shall now analyze the effective
potential in order to extract information about the particle
dynamics. Introducing an $l$-dependent effective mass
\be
\tilde{m}^2 = m^2 + \frac{l^{2}}{\beta},
\label{meff}
\ee
we can rewrite $V_{eff}$ in the form,
\begin{equation}
2m V_{eff} = \left(1-\frac{2M}{r}\right) \left[ \tilde{m}^2 +
\frac{l^{2}}{ r^{2}} \right].
\label{v1}
\end{equation}

\begin{figure}[htb]
{\centerline{\includegraphics[width=10cm, height=6cm] {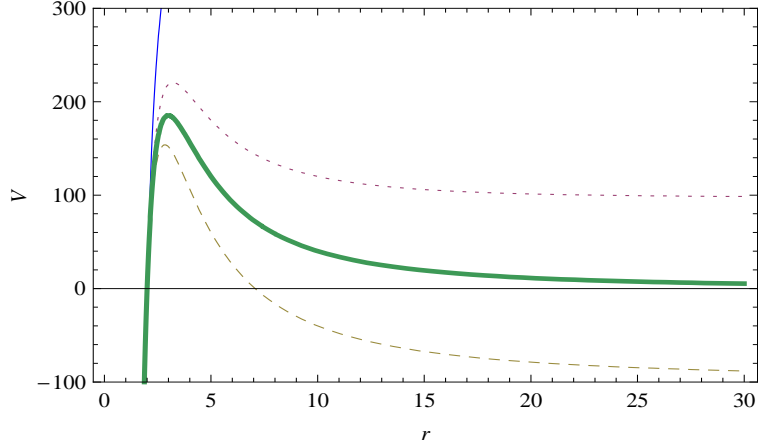}}}
\caption{Plot for the variation of the effective potential with
$r$ for $l=100$ with different values of $\beta$}
\label{fig1}
\end{figure}

\n
The effect of different values of $\beta$ on the dynamics of a particle with
mass $m$ can be seen in Figure 1, where we plot $V(r)~vs.~r$ for different
$\beta$ at a fixed $l=100$. In this figure the normal line corresponds
to $\beta =10$ whereas the dotted, thick and dashed lines represent the values
$\beta = 50, 10^{5}, -50$ respectively. Also, we have taken $M =1$, $m =0.1$.
We now discuss the various cases in some more detail.
\vspace{2ex}

\n {\it\textbf{{Positive}} $\beta $:}\\ The $\beta \rightarrow
\infty$ limit, where the conventional particle is recovered, is
given by the $\beta = 10^{5}$ plot. On the other hand, the
potential diverges for a small value of $\beta$ (indicated by
normal line) so that one cannot have a  bound state for a particle
in the small $\beta$ limit. This happens due to a huge amount of
the binding energy required from the additional amount of mass
term given by $l^2/\beta$.  In order to have a stable orbit for
the particle, the effective potential $V(r)$ should have a
minimum; so we must have

\begin{equation}
\frac{dV}{dr} = 0.
\end{equation}
The above relation gives us a quadratic equation to solve and the
equation is the following:

\begin{equation}
M \tilde m^{2} r^{2} - l^{2} r + 3 M l^{2} = 0.
\end{equation}
The two solutions of the above equation are given by:
\begin{equation}
r_\pm = \frac{l^2 \pm \sqrt{l^4-12 M^2 l^2(m^2+\frac{l^2}{\beta})}}
 {2 M(m^2+\frac{l^2}{\beta})}.
\end{equation}
One of these two solutions represents a maximum, i.e. an unstable orbit
and the other represents a minimum, i.e. a
stable orbit. The stability condition turns out to be,
\begin{equation}
l^2 \geq \frac{12 M^2 m^2 \beta}{\beta-12 M^2}.
\end{equation}
{\it\textbf{{Negative}} $\beta $:}\\However, strikingly new behavior
appears for negative $\beta $ with $l^2 < | \bg | m^2$. This
is seen in the Figure 1 for $\beta =-50$. The effective potential
changes sign after some value of $r$,
indicating a change in the behavior of the particle trajectory at some point.
The cosmological implications of this sign-flip will be discussed in
the next section.

\section{Cosmological implications of negative $\beta$}

So far we have analyzed the significance of the dynamics of the
generalized particle. In this section we shall point out a crucial
implication of the scenario in the cosmological context. From
embedding geometry using the Gauss-Codazzi equation \cite{poisson}
it can be shown that a $(D-1)$-dimensional surface representing a
cosmological Friedmann-Robertson-Walker (FRW) metric can be
embedded consistently in a $D$-dimensional black hole space-time
in such a way that the expansion of the FRW surface is realized by
the particle trajectory along the radial direction in the
gravitational field of the black hole \cite{bwg,
embed1,embed2,embed,pal}. In the present article, the particle
motion represents a 3-dimensional FRW metric and the effective
potential contains essential information for the evolution of the
embedded cosmological surface. Below we shall show how this can be
realized, followed by a discussion of its cosmological
implications.

Let us take a careful look at the expression for the effective
potential (\ref{v1}). It contains additional terms arising from
$\beta$ in the generalized particle approach. It is straightforward
to verify that the scenario is identical to the motion of a point
mass in the field of a static, spherically symmetric metric
\be ds_4^2 = - F(r) dt^2 + \frac{dr^2}{F(r)} +
r^2 d\Omega_2^2 \label{effmet} \ee
where
\be d\Omega_2^2 =
\frac{d\sg^{2}}{1-k\sg^{2}} + \sg^{2} d\phi^{2},
\ee
provided the metric function $F(r)$ is chosen as
\be
F(r) = \left( 1 - \frac{2M}{r} \right) \left(
 \frac{\tilde{m}^2 r^2 + l^2}{m^2 r^2+ l^2} \right)
\label{fr} \ee The geodesic equations associated with this metric
(\ref{effmet}) reproduce precisely the same effective potential as
that in eq.\ (\ref{v1}). How the above form of the metric can be
derived directly from solving the Einstein equations with some
specific source is presently an open problem. The reader may
however note that this effective metric falls in the class of
several complicated spherically symmetric metrics available in the
literature \cite{chandra} (Some more exotic metrics are listed in
\cite{ssssol}); here the only point is that this effective metric
arises from the generalized particle dynamics. In reality, of
course, the background space-time still remains of the
Schwarzschild type.

Our intention is, rather, to embed a 3-dimensional FRW metric \be
ds_3^2 = - d\tau^2 + a^2(\tau) d\Omega_2^2 \ee (where $d\Omega_2^2$ is
the two-space representing the flat, open or closed spaces) into
the space-time given by eq.\ (\ref{effmet}). With the effective
metric function (\ref{fr}), and identifying the scale factor with
the radial trajectory so that the expansion of the cosmological
universe is realized by the radial motion of the particle, the
tangents (4-velocity) and normals to the surface satisfying the
orthonormality and normalization conditions are given by \bea
u^\mu &\equiv& \left( \frac{\sqrt{F(a) + \dot a^2}}{F(a)}, ~\dot
a, ~0, ~0
\right) \nonumber \\
n^\mu &\equiv& \left( -\frac{\dot a}{F(a)}, ~ - \sqrt{F(a) + \dot a^2}, ~0, ~0
 \right)
\eea
Here, and throughout the rest of the discussion, we identify $r(\tau)$
with the scale factor $a(\tau)$.

Further, the extrinsic curvature turns out to be
\be
K_{ij} = \frac{\sqrt{F(a) + \dot a^2}}{a} ~\tilde g_{ij} ~~; ~~
K_{\tau\tau} = \frac{d}{da}  \left(\sqrt{F(a) + \dot a^2} \right)
\label{extcurv}
\ee
where $\tilde g_{\mu \nu}$ is the induced metric of the 3-dimensional
FRW surface.

The junction conditions, along with $Z_2$-symmetry, relates the
extrinsic curvature to the effective surface stress-energy tensor
$S_{\mu\nu}$ by
\be
K_{\mu\nu} = - 8 \pi G
\left(S_{\mu\nu} - \frac{1}{3} S \tilde g_{\mu \nu} \right)
\ee
With the extrinsic curvature (\ref{extcurv}), the square of the
above equation, immediately leads to
\be
\left(\frac{\dot a}{a} \right)^2 = -\frac{F(a)}{a^2} + \frac{8 \pi G}{3} \rho
\ee
where $\rho$ is the effective surface density that arises from matter
and the surface tension. Written explicitly in terms of the metric
function (\ref{fr}), we have the Friedmann equation on the
cosmological surface
\be
\left(\frac{\dot a}{a} \right)^2 =
\frac{8 \pi G}{3} \rho - \frac{k}{a^2} + \frac{2M}{a^3} -
\frac{l^2}{\beta} \left(\frac{1-2M/a}{l^2+m^2a^2} \right)
\label{fried}
\ee
The effective matter conservation
equation on the surface holds good. Consequently, we arrive at the
following Raychaudhuri equation for expansion
\be
\frac{\ddot a}{a} = - \frac{4 \pi G}{3} (\rho + 3p)
- \frac{M}{a^3} + T(\bg),
\label{rc}
\ee
where the $\bg$-dependent terms are
\be
T(\beta) = -\frac{l^2}{\beta (l^2+ m^2 a^2)^2}
 \left[l^2 -\frac{Ml^2}{a} + Mm^2 a \right].
\ee
The evolution of the 3-dimensional cosmological universe is
governed by the above three equations.  Eq.\ (\ref{rc}) needs special
attention in this regard. The $M/a^3$ term is a radiation-like
effect (note the cosmology is now 3-dimensional) from the
Weyl tensor of the black hole and hence is negligible for late
time evolution. This is analogous to the dark radiation in a
braneworld context \cite{bwg}. The terms $T(\bg)$, however, are
not so trivial, and in fact these terms give rise to the most
important physical conclusion.

Indeed, the above correction to the evolution equations (\ref{rc})
of the embedded cosmological surface is quite significant when the
$\beta$ parameter takes negative value. With a negative $\beta$,
the term outside the square bracket is positive definite.
Consequently, this correction term $T(\beta)$ has a positive
contribution to the expansion equation (\ref{rc}) for the relevant
region $a > 2M$ (the particle trajectory is outside the black hole
horizon), thereby resulting in an effective negative pressure
which becomes significant at late time. This term essentially
leads to late time accelerating phase.  This is clearly
demonstrated in {\bf{Figure 1}}. The model is thus capable of
explaining a transition from the decelerating to accelerating
phases of the universe with an effective negative pressure arising
from a negative $\beta$. Thus the framework has the potentiality
to provide a source for dark energy.

Our claim can further be established from the analysis of deceleration parameter
which is one of the major observable quantities for the present universe.
Considering the terms containing $\beta$ as the driving force for the acceleration
of the universe (which implies that this is the dominant contribution to cosmic density
at late time) and neglecting the effect of the matter sector, the deceleration parameter
turns out to be

\begin{equation}
q= -\frac{\ddot{a}/a}{(\dot{a}/a)^2} =  -\frac{l^2}{l^2+m^2 a^2}-\frac{M}{a-2M}.
\end{equation}

This clearly reveals that the deceleration parameter is negative
for the cosmologically relevant region $a > 2M$, confirming an
accelerated expansion at late time. This behavior is further
transparent from {\bf{Figure 2}} which shows the variation of the
deceleration parameter $q$ with Hubble parameter $H= \dot a/a$ for
the representative values of the constants.

\begin{figure}[htb]
{\centerline{\includegraphics[width=8cm, height=5cm] {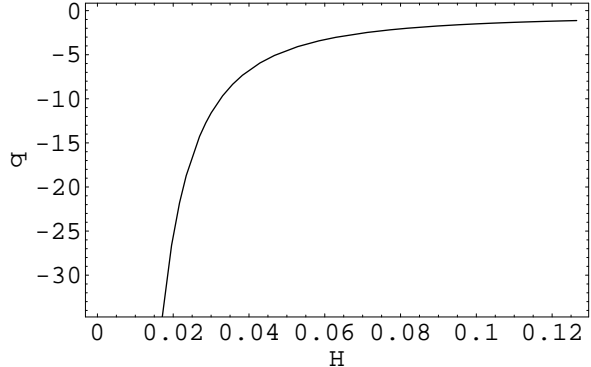}}}
\caption{Variation of the deceleration parameter $q$ with
 Hubble parameter $H$ for $l=100$, $\beta=-50$, $M=1$, $m=0.1$}
\end{figure}

The behavior for the scale factor can also be analyzed to some
extent with the above considerations of $\beta$-dominance. In the
large $l$ and small $m$ limit, an approximate solution for the
Friedmann equations can be obtained as

\begin{equation}
a(t) \approx M + \frac{1}{2} \left[e^{t/\sqrt{-\beta}} -M^2 e^{-t/\sqrt{-\beta}} \right]
\end{equation}
As already mentioned, $\beta$ is negative here. In the above
equation, the first term is nothing but a scaling and with $M=1$
the second term reduces to $\sinh (t/\sqrt{-\beta})$, which
interpolates between a decelerating phase (matter-dominated) at
early time and an accelerated expansion at late time. Thus our
model behaves pretty close to $\Lambda$CDM, with the inverse of
the $\beta$ parameter playing the role of the cosmological
constant. Because of this identification, it is worthwhile to
emphasize that large negative $\beta $ is consistent with small
positive value for $\Lambda $ that is favored observationally.
This makes the scenario further interesting from observational
ground since now one can calculate the other observable parameters
and compare them with highly accurate observational data.

From the particle dynamics point of view, an effective negative
pressure from a negative $\beta$ is, indeed, what is expected.
This will be transparent from the expression of the effective mass
given in the last section by $\tilde m^2 = m^2 +
\frac{l^2}{\beta}$. Clearly, a negative value for the $\beta$
parameter will result in a reduced effective mass for the particle
as compared to its physical mass $(\tilde m^2 < m^2)$. This
apparent loss in mass will make the particle feel as if it is
being acted upon by a repulsive force, which is concretized by an
effective negative pressure in the Friedmann equations
(\ref{fried}) and (\ref{rc}), thereby leading to a
late-accelerating behavior for the cosmological surface from our
analysis. The sign-flip in the effective potential in {\bf{Figure
1}} for negative $\beta$ shows this behavior.

This gives a major cosmological implication of the paradigm. One
can readily see that the computations are literally the same for a
4-dimensional cosmological metric embedded in a 5-dimensional
black hole space-time. Thus, there is absolutely no problem in
applying the model to real-life cosmological analysis. A thorough
investigation in this direction will be reported in future.


\section{Summary and outlook}

In this article we have developed a generalized framework for
dynamics of a particle moving in a fixed background geometry,
based on a reparametrization-invariant action containing the
information of the symmetry properties of the space-time. In our
analysis, we have considered hamiltonian as well as lagrangian
formalisms and have shown that both of them lead to same results,
thereby confirming the consistency of the theory.

Applying our framework to the case of Schwarzschild geometry, we
have further analyzed the effective potential as usual in GR. Our
analysis reveals some distinct features of the model. First, the
physical mass of the particle is replaced by a more relevant
effective mass which is a sum-total of the physical mass and an
additional mass-like effect arising from the term $\beta$
introduced in the theory. Secondly, using rigorous analysis with
different plots for the variation of the effective potential with
radial distance, we find that among a spectrum of possibilities
for different value of the $\beta$ parameter, the usual particle
dynamics for Schwarzschild space-time is recovered only for the
$\beta \rightarrow \infty$ limit. Further, the effective mass
results in an additional binding energy for the particle so that a
stable orbit is now obtained for a larger value of the angular
momentum. A last and significant outcome of our analysis is that
we uncover a source for negative pressure analogous to dark energy
for a cosmological metric moving along the particle trajectory.

Since throughout the effective potential analysis, the physical
mass is replaced by the effective mass the quantitative
estimations for a massive particle are going to be different from
the GR counterparts. One can estimate the quantities and subject
them to experimental verifications. Also, the action we choose is
one of the simplest extension of the single-particle action that
preserves reparametrization-invariance. More general actions can
lead to more dramatic results. For example, a suitable choice of
the action may lead to a noncommutative structure which may
further be utilized to verify whether the space-time becomes
noncommutative in the vicinity of the black hole horizon.

A thorough investigation of the evolution of the universe
providing a late-accelerating phase with the modified Friedmann
equations is one of the major open issues. At the very first
point, it is interesting to look for any analytical or numerical
solution for the scale factor from equations (\ref{fried}) and
(\ref{rc}), which will explicitly show the late-accelerating
behavior. Secondly, one can calculate different observable
quantities \cite{de}, such as the deceleration parameter $q$, the
Hubble parameter $H(z)$, age of the universe $t_{0}$, luminosity
distance $d_L$, statefinder parameters $\{r, s\}$
\cite{statefinder}, $Om(z)$ parameter, acceleration probe $\bar q$
\cite{newprobe} and the other observable quantities as well. The
next step is to  confront them with observations. The
next step is to  confront them with observations.
Some of the issues have been discussed to some extent in the present article.
We hope to address some of these issues in near future.

\section*{Acknowledgments}

SP thanks Sayan Kar for illuminating discussions on Killing vectors.


\begin{thebibliography}{99}
\bibitem{regge} A.J.Hanson, T.Regge and C.Teitelboim, {\it Constrained
Hamiltonian System}, Roma, Accademia Nazionale Dei  Lincei, (1976);
S.Ghosh, Phys.Lett. B338 (1994) 235 (arXiv:hep-th/9406089);
Erratum-ibid. B347 (1995) 468
\bibitem{rvh} R.H.\ Rietdijk and J.W.\ van Holten, Class.\ Quantum Grav.\ 7
(1990), 247; J.W.\ van Holten, {\em Relativistic Dynamics of Spin in
Strong External Fields}, Proc.\ 4th Hellenic School on Elementary
Particle Physics, Vol.\ II; Eds.\ E.N.\ Gazis, G.\ Koutsoumbas,
N.D.\ Tracas and G.\ Zoupanos (Corfu, 1992)
\bibitem{pioneer} S.G.\ Turyshev et al., Int.\ J.\ Mod.\ Phys.\ D15 (2006), 1
\bibitem{holt} For studies on Killing vectors, directly relevant for the present work, see .
J.W. van Holten,  Phys.Rev.D75 (2007) 025027  (hep-th/0612216);
R.H. Rietdijk and J.W. van Holten, Nucl.Phys.B472 (1996) 427
(hep-th/9511166); R.H. Rietdijk  and J.W. van Holten,
Nucl.Phys.B404 (1993) 42,.(hep-th/9303112); Symmetries and motions
in manifolds. J.W. van Holten and R.H. Rietdijk,
J.Geom.Phys.11:559,1993 (hep-th/9205074).
\bibitem{bwg} R. Maartens, Living Rev. Relativity 7 (2004) 7.
\bibitem{embed1}M.Gogberashvili, Europhys.Lett. 49 (2000) 396-399
 (arXiv:hep-ph/9812365v1).
\bibitem{embed2}C. Barcelo and M. Visser, Phys. Lett.  B {\bf 482}, 183 (2000).
\bibitem{dir}P. A. M. Dirac, {\em Lectures on Quantum Mechanics},
Yeshiva University Press, New York, 1964.
\bibitem{birkhoff} G. D. Birkhoff, {\em Relativity and Modern Physics},
Harvard University
Press, Cambridge (1923).  For a proof of generalised Birkhoff's theorem,
see R. D'Inverno,
{\em Introducing Einstein's Relativity}, Oxford University  Press,
Oxford (1995).
\bibitem{brinketal} L.\ Brink. P.\ Di Vecchia and P.\ Howe,
 Nucl.\ Phys.\ B118 (1977), 76
\bibitem{bks}R. Banerjee, S. Kulkarni and S. Samanta, JHEP {\bf
05} (2006) 077 (arXiv:hep-th/0602151).
\bibitem{snyder}H. S. Snyder, Phys. Rev. {\bf 71} (1947) 38.
\bibitem{carroll} S. M. Carroll, {\em Lecture Notes on General Relativity},
gr-qc/9712019
\bibitem{poisson} E. Poisson, {\em A Relativist's Toolkit:
The Mathematics of Black-Hole Mechanics}, Cambridge University Press,
Cambridge (2004);  S. Chandrasekhar,  {\em The Mathematical Theory of
Black Holes}, Clarendon Press, Oxford (1992).
\bibitem{embed}
J. Garriga and M. Sasaki, Phys. Rev. {\bf D62} (2000)  043523;
P. Kraus, JHEP {\bf 12}, 011 (1999);
D. Ida, JHEP {\bf 09}, 014 (2000);
S. Mukohyama, T. Shiromizu and K Maeda, Phys. Rev. D {\bf 62}, 024028 (2000);
P. Bowcock, C. Charmousis and R. Gregory, Class. Quant. Grav. {\bf 17},
 4745 (2000).
\bibitem{pal} S. Pal, Phys. Rev. {\bf D 74}, 024005 (2006);
S. Pal, Phys. Rev. {\bf D 78}, 043517 (2008).
\bibitem{chandra} S. Chandrasekhar,  {\em The Mathematical Theory of Black Holes},
Clarendon Press, Oxford (1992).
\bibitem{ssssol} P. C. Vaidya, Nature {\bf 171} (1953) 260;
C. W. Misner, Phys. Rev. {\bf 137} (1965) B1360;
R. W. Lindquist, R. A. Schwartz and C. W. Misner, Phys. Rev. {\bf 137} (1965) B1364;
V. V. Narlikar, G. K. Patwardhan and P. C. Vaidya, Proc. Natl. Inst. Sci. India
{\bf 9}, 229 (1943);
B. Kuchowicz, Acta Phys. Pol. {\bf B 3}, 209 (1972);
 H. A. Buchdahl, Astrophys. J. {\bf 140}, 1512 (1964);
A. Chamblin, S. W. Hawking and H. S. Reall, Phys. Rev. {\bf D 61}, 065007 (2000);
C. Germani and R. Maartens, Phys. Rev. {\bf D 64} (2001) 124010;
R. Casadio, A. Fabbri and L. Mazzacurati, Phys. Rev. {\bf D 65} (2002) 084040;
P. Kanti and K. Tamvakis, Phys. Rev. {\bf D 65}, 084010 (2002);
S. Shankaranarayanan and N. Dadhich, Int. J. Mod. Phys. {\bf D 13} (2004) 1095;
A.  N. Aliev and  A. E. Gumrukcuoglu, Phys. Rev. {\bf D71} (2005) 104027;
A. S. Majumdar and N. Mukherjee, Int. J. Mod. Phys. {\bf D14} (2005) 1095;
S. Creek, R. Gregory, P. Kanti and B. Mistry, Class. Quant. Grav. {\bf 23}, 6633 (2006);
 A. L. Fitzpatrick, L. Randall and T. Wiseman, JHEP {\bf 0611}, 033 (2006);
R. Gregory, Lect. Notes Phys. {\bf 769}, 259 (2009) [arXiv:0804.2595].
\bibitem{de} V. Sahni, Lect. Notes Phys. \textbf{653} (2004) 141;
E. J. Copeland, M. Sami and S. Tsujikawa, Int. J. Mod. Phys. {\bf D 15},
 1753 (2006);
J. Frieman, M. Turner  and D. Huterer, arXiv:0803.0982 [astro-ph]
\bibitem{statefinder} V. Sahni, T. D. Saini, A. A. Starobinsky and U. Alam,
JETP Lett. {\bf 77} (2003) 201; Pisma Zh. Eksp. Teor. Fiz. {\bf 77} (2003) 249;
U. Alam, V. Sahni, T. D. Saini and A. A. Starobinsky,
Mon. Not. Roy. Astron. Soc. {\bf 344} (2003) 1057
\bibitem{newprobe} V. Sahni, A. Shafieloo and A. A. Starobinsky,
Phys. Rev. {\bf D 78}, 103502 (2008)
[arXiv: 0807.3548]





\end{thebibliography}
\end{document}